\documentclass[preprint,12pt]{elsarticle}
\usepackage{amssymb}
\usepackage{amsmath}
\usepackage{epsfig}
\usepackage{soul}
\usepackage{geometry}

\newcommand{\IC}{infective contact}

\begin{document}

\def\IC{infective contact}
\def\ROO{$\alpha$}
\def\ROOenfla{\alpha}
\def\ROd{R_{0}^{(d)}}
\def\RO{$R_0$}
\def\RR{X_{rr}}
\def\IR{X_{ir}}
  \def\II{X_{ii}}  
  \def\SR{X_{sr}}
 \def\SI{X_{si}}
 \def\SII{(X_{si})^2}
 \def\SS{X_{ss}}
 \def\S{X_{s}}
  \def\I{X_{i}}
   \def\R{X_{r}}
   \def\NRR{N_{RR}^{(d)}}
\def\NIR{N_{IR}^{(d)}}
  \def\NII{N_{II}^ {(d)}}  
  \def\NSR{N_{SR}^{(d)}}
 \def\NSI{N_{SI}^{(d)}}
 \def\NSS{N_{SS}^{(d)}}
 \def\NS{N_{S}^{(d)}}
  \def\NI{N_{I}^{(d)}}
   \def\NR{N_{R}^{(d)}}
   \def\EL{\scriptscriptstyle L}
   \def\EG{\scriptscriptstyle G}
\vskip 0.5cm
\begin{frontmatter}
  \title{
  Exploring the threshold of epidemic spreading
 for a stochastic SIR model with local and global contacts
}
\author[add1]{Gabriel Fabricius}
\ead{fabricius@fisica.unlp.edu.ar}
\author[add2]{Alberto Maltz}
\ead{alberto@mate.unlp.edu.ar}

\address[add1]{Instituto de Investigaciones Fisicoqu\'{\i}ımicas Te\'oricas y Aplicadas,
	Facultad de Ciencias Exactas, Universidad Nacional de La Plata,
	CC 16, Suc. 4, 1900 La Plata, Argentina}
\address[add2]{Departamento de Matem\'atica,
	Facultad de Ciencias Exactas, Universidad Nacional de La Plata,
	CC 72, Correo Central, 1900 La Plata, Argentina}
\begin{abstract}
The spread of an epidemic process is considered 
 in the context of a spatial SIR 
stochastic model that includes a parameter $0\le p\le 1$  that assigns 
weights $p$ and $1- p$ to global and local infective contacts respectively.
The model was previously studied by other authors in different contexts.
In this work we characterized the behavior of the system around the threshold
for epidemic spreading.
We first used a  deterministic approximation of the stochastic model 
 and checked the existence of a 
threshold value of $p$ for exponential epidemic spread.  
An analytical expression,  which defines  a function of the quotient $\alpha$ 
between the transmission and recovery rates, 
is obtained to approximate this threshold. 
We then performed different analyses based on intensive stochastic 
simulations and found that this expression is also a good estimate 
for a similar 
threshold value of $p$ obtained in the stochastic model. 
The dynamics of the average number of infected individuals and the 
average size of outbreaks 
show a behavior across the threshold that is well described 
by the deterministic approximation.
The distributions of the outbreak sizes at the threshold
present common features for all the cases considered  corresponding to different values of $\alpha>1$.
These features are otherwise already known to hold for the standard
 stochastic SIR model at its threshold, $\alpha=1$:
(i) the probability of having an outbreak of size $n$ goes asymptotically as $n^{-3/2}$
for an infinite system, (ii) the maximal size of an outbreak scales as $N^{2/3}$ for a finite
system of size $N$. 

\end{abstract}
\end{frontmatter}

\section{Introduction}

The SIR model is probably the most widely used model in mathematical epidemiology \cite{bailey, anderson, libroKR}.
Since W. Kermack and A. McKendrick employed it to describe the development of a plague epidemic
in Bombay \cite{kermack}, it has been extensively used for  many purposes.
Sometimes as the kernel of more complex epidemiological models in the field of infectious disease transmission \cite{anderson, heester2}
and other times to study spreading phenomena in other fileds such as rumor propagation \cite{daley}, 
computer viruses \cite{mishra}, information diffusion in Web forums \cite{woo} or investors' behavior in stock markets. \cite{shive}.

 In the deterministic version of the model, the three epidemiological 
 classes are represented by three continuous variables 
 (S: susceptible, I: infected, and R: recovered) that evolve in time 
 according to a set of ordinary coupled differential equations. 
 It is easy to treat mathematically, has a straightforward interpretation 
 and leads to simple and important predictions \cite{kermack}. 
 However, from the very beginning it was discovered that for some applications,  
 it was necessary to consider the fluctuations and the discrete nature of 
 the population and of the processes  involved (infection and recovery) 
  \cite{bailey,bartlett56}. For example,  using a stochastic 
 version of the SIR model, M. S. Bartlett showed that stochasticity was an essential 
 aspect to be considered to explain the persistence of measles as a 
 function of the city size in several cities of England and Wales 
 \cite{bartlett56,bartlett57}. 

The threshold for epidemic spreading 
that in both versions of the SIR model (deterministic and stochastic) occurs when
the transmission rate ($\beta$) equals the recovery rate ($\gamma$) and so,  
the quotient $\alpha \equiv \beta / \gamma=1$ is of particular interest. 
The properties of the stochastic SIR model at the threshold 
 have been extensively studied lately by rigorous mathematical theory, more empirical treatments, and computer simulations
\cite{martinLof1998,bennaim2004,kessler2007,bennaim2012}.

In the present work we study a stochastic SIR model on the lattice with local and
global contacts where the weight of
the global contacts is taken into account through a single parameter $p$.
It is inspired in Watts-Strogatz model \cite{watts-strogatz} 
and has already been used by other authors to study transmission of diseases of high 
\ROO\ values  \cite{verdasca,simoes,dottori,maltz}.
The purpose of using these simple models in this field is  to gain   insight
into the qualitative trends observed when the global contacts are lowered. To 
obtain an accurate description of the transmission process, a more realistic
network well suited to represent human interactions should be used.
On the other hand,
SIR type models that combine local contacts in a square lattice with some kind of global connection 
have been proposed to study the spread of infectious diseases in plants and animals \cite{gilligan,liccardo,tischendorf}.
  In these cases,
 where the lattice  could eventually be a good enough
representation of the ``spatiality" of the problem,
predictions could also have a quantitative character.

In this work we study  the model for low \ROO\ values
since we are interested in the behavior  around the threshold
for epidemic spreading. 
In models where the hypothesis of uniform mixing holds,
the threshold is given by the condition $R_0$=1, where 
$R_0$ is the basic reproductive number, defined as
the average number of secondary infections produced by one infected 
individual in a completely susceptible population \cite{anderson}.
This is the case of the classic SIR model where $R_0=\alpha$.
In models where uniform mixing is broken, as in the one studied in the present work,
it is known that the concept of $R_0$ is meaningless \cite{anderson,failurer0,riley}, and so,
 the threshold condition is something to be explored.
We here show that it is possible to define a threshold for exponential spreading that 
in this model (with two free parameters) is not a point as in the classical SIR model but 
a curve in the $(p, \alpha)$ plane.
We found that for any $\alpha>1$ considered, 
by sufficiently lowering  the value of $p$, 
the threshold can be crossed
 producing a drastic reduction of the probability for a major outbreak.
We managed to characterize several model properties
around the threshold through the analysys of stochastic simulations
and by using a deterministic approximation developed in a previous work \cite{maltz}.

The work is organized as follows: we first introduce 
 the stochastic model (SM) and the deterministic approximation (DA).
Then, using the DA, we can build a phase diagram in the parameter space detecting
the region in which it is possible to prevent the 
exponential spread. We perform intensive stochastic simulations 
and find that average magnitudes follow the same trends
predicted by the DA. In particular the dynamical behavior of the average number of infected individuals
makes it possible to  unambiguously define 
the threshold for exponential spreading in the SM 
for all the values of $\alpha$ considered.
We then analyze the distributions of the outbreak sizes at the epidemic threshold
and obtain several features that are known to hold for the classical SIR model. A summary of our findings
is finally given in the conclusions.

\section{The stochastic model and the deterministic approximation}

\subsection{The stochastic model}
We consider the stochastic model studied in \cite{dottori}, 
but since in the present work we focus on the epidemic spread
only, we do not consider the births and deaths. The population consists
of $N$ individuals
identified with the sites of an $L \times L$ square lattice  
with periodic boundary conditions. They may 
be in one of the three epidemiological states: $S$, $I$ or $R$ 
(susceptible, infected or recovered). The dynamics of the model are described by a 
stochastic Markovian process in which an individual 
may experience one of these two changes in its state:
$S\rightarrow I$ (infection) or $I\rightarrow R$ (recovery).
Infections occur through infective contacts among susceptible and infected individuals. 
We define an infective contact as a contact between two
individuals such that if one individual is susceptible and the other infected, the former becomes infected. We assume that
an individual at a given site  has an \IC\ with a randomly chosen individual on the lattice with transition  rate  $p \beta$, and with one of its four nearest neighbors (also randomly chosen) with transition  rate $(1-p) \beta$.
By changing $p$, we may change 
the relative weight of the global and local
contacts in the system. The case $p$=1 corresponds to the classical 
SIR model (uniform mixing) where an individual may contact 
any other individual in the system.
On the other hand, the case $p$=0 corresponds to the square lattice where an individual may only contact one of its   four nearest neighbors. Recovery from infection in this model is the same for every site and occurs at a transition  rate $\gamma$.

The state of the system, $\Gamma=(E_1, E_2, ...,E_N)$, is defined by specifying
$E_j$ (the state of site $j$) for the $N$ sites of the lattice.
Three variables of interest are the number of individuals
in the system that are in state $S$, $I$ and $R$ that we
call: $N_S$, $N_I$, and $N_R$ respectively.

The probability transition rates for infection and recovery processes 
at site ``$j$" are
\begin{flalign}
&W_{inf}^j=\left[ p \ \beta \ \frac{N_I}{N} + (1-p) \beta \ 
            \frac{1}{4} \sum_{j'\in \nu_j} 
            \delta _{E_{j'},I} \right] \delta _{E_{j},S} \nonumber \\  
&W_{rec}^j=\gamma \delta _{E_{j},I}   \nonumber
\end{flalign}
where $\delta_{A B}$ is one if states $A$ and $B$ are the same, 
and zero if not. Index
$j'$ in the sum runs over the $4$ neighbors of site $j$ (we call 
this set of sites $\nu_j$).

Stochastic simulations are performed using the Gillespie
algorithm \cite{gillespie}. 
Each simulation begins in the same initial state where $N-1$ individuals
are susceptible and one individual is infected.
A Markov chain 
\[
\Gamma (t_1) \rightarrow \Gamma (t_2) \rightarrow ...\Gamma (t_{ext}) 
\]
with a  set of exponentially distributed times $t_1$, $t_2$, ...$t_{ext}$ is generated,
where $t_{ext}$ (the time at extinction)  is the first time with $N_I=0$.
$\Gamma (t_{ext})$ is an absorbing state and $N_R(t_{ext})$ (the number
of recovered individuals) is the total number of individuals
that have experienced the infection during the dynamic evolution
of the disease from the initial state to its extinction. We also refer to
$N_R(t_{ext})$ as the "outbreak size".

For  most of the calculations in this work 
we take $L=800$, $N=N_0=640,000$, but in Section \ref{sectionthreshold}
 we also consider
$N=2N_0$ and $N=4N_0$ to study size effects.

\subsection{Deterministic approximation}

In the present work  we  use a  deterministic approximation of
the stochastic model in \cite{dottori}. In this approximation,
 developed in \cite{maltz},
the local infective contacts are treated using a pair 
approximation scheme with second moment closure.

          In the deterministic context we denote $\NS$, $\NI$, $\NR$, the time
          functions whose values are  the number of individuals of each type, and
            $\NSS$, $\NSI$, $\NSR$, $\NII$, $\NIR$, $\NRR$, the similar ones
          for  the number of  pairs 
          formed by two  neighboring individuals whose type corresponds to the 
         subscripts; for example, $\NSR$ represents the number of
           pairs formed by a susceptible and a recovered neighboring individual.
           
          Then we define the nine functions $\I=\NI/N$, $\S=\NS/N$, $\R=\NR/N$, 
           $\SS=\NSS/N$, $\SI=\NSI/N$, $\SR= \NSR/N$,
         \   $\II=\NII/N$, $\IR=\NIR/N$, $\RR=\NRR/N$.
           
           In \cite{maltz} we construct a system of nine 
           differential equations having these nine unknowns. Then, using several relationships
           between the unknowns, the system is  reduced to five equations and
           the unknowns $\S$, $\I$ ,$\SS$, $\SI$, $\II$.
In the present case,
where birth and death are not considered, the five equations of the DA are:

 \begin{align}
 \frac{d\S}{dt} =& -p\beta\S\I-\frac{(1-p)\beta\SI}{4}    \label{ecu1}  \\ \nonumber \\
 \frac{d\I}{dt} =& -\gamma\I+p\beta\S\I+\frac{(1-p)\beta\SI}{4}  \label{ecu2} \\ \nonumber \\
 \frac{d\SS}{dt} =& -2p\beta\I\SS-\frac{3(1-p)\beta\SI\SS}{8\S} \label{ecu3}  \\ \nonumber \\
  \frac{d\SI}{dt} =& -p\beta\I\SI-3(1-p)\beta\left.{\bigg{(}}\frac{\SII}{16\S}-\frac{\SI\SS}{8\S}\right.{\bigg{)}} \nonumber \\ \nonumber \\
                   & +2p\beta\I\SS  - \left.{\bigg{(}}\frac{(1-p)\beta}{4}+\gamma\right.{\bigg{)}}\SI \label{ecu4}  \\ \nonumber \\
 \frac{d\II}{dt} =& -2\gamma\II+p\beta\I\SI+(1-p)\beta\left.{\bigg{(}}\frac{3\SII}{16\S}+\frac{\SI}{4}\right.{\bigg{)}} \label{ecu5}
 \end{align}

We take $\S(0)=(N-1)/N, \  \I(0)= 1/N, \ \SS(0)= (2N-4)/N ,\  \SI(0)= 4/N,\  \II(0)=0$ as initial values,  which correspond  to  the presence
of only one infected individual, and $N-1$ susceptible ones. 

 Now we define $\ROd$, which we will call ``$R_0$ of the deterministic 
approximation".
 Consider a fixed time $t$. By equation (\ref{ecu2}) the amount of new infective cases  in the (infinitesimal)
time interval between $t$ and $t+dt$ is  $N\beta\big{(}{(1-p)\SI(t)}/{4}+ p\S(t)\I(t) \big{)}dt$. 
   If the only initial infected individual remains in this state at time $t$,
   we estimate  the number of new   
  infections generated between times $t$ and $t+dt$ by this individual 
as the quotient between that amount and
$\NI(t)=N\I(t)=$ number of infected individuals at time $t$.
   To take into account the recovery possibility of this individual, we multiply by the ``probabilistic''  factor $e^{-\gamma t}$, thus
    arriving at
\begin{equation}  
 \ROd=  \int_{0}^{\infty}\beta\bigg{(}\frac{(1-p)\SI(t)}{4\I(t)}+p\S(t)\bigg{)} e^{-\gamma t}dt.  \label{eqnr0det} 
\end{equation}
   As expected, for $p=1$ we obtain the value \ROO\ corresponding to the SIR model (see \cite{heester1}):  
  
\begin{equation}  
    \int_{0}^{\infty}\beta \S(t)e^{-\gamma t}dt =\frac{\beta}{\gamma}= \ROOenfla \label{HDeqn}.
\end{equation}  
               
   When necessary, we will calculate $\ROd$ numerically, for different values of
 $p$, simultaneously with the resolution of the system, 
 which is performed
using the Euler's Method with a time step of  0.01 day. 

\section{Results and Discussion}

Henceforth, we take $1/\gamma$ (the mean duration of infection) as the unit of time. 
This is equivalent to making the change of variables: $\tau = \gamma t$. 
Doing this  in  equations (\ref{ecu1}) to (\ref{HDeqn}) is equivalent
to substituting $\gamma$ by 1, $\beta$ by $\beta/\gamma$ and $t$ by $\tau$.
Since $\beta/\gamma$=\ROO, there are only two free parameters 
in our study: \ROO\ and $p$.
In this work we focus on the case $0 \leq p \leq 1$ and $1 \leq$ \ROO $\leq 2$.

\subsection{Deterministic approximation predictions}

In this section we explore the dynamical behavior of the DA system
for low \ROO\ values.

\subsubsection{Condition for exponential epidemic spread }

For $p=1$ the DA reduces to the SIR model where it is  well known that  $\I(\tau)$ behaves as
\[
\I(\tau) \simeq \I(0) e^{(\ROOenfla - 1) \tau}
\]
as long as $\S(\tau)$ does not fall appreciably from $1$.
Then the condition \ROO$>1$ determines the exponential spread of the epidemic in this case.

For $p <1$, in our previous work \cite{maltz}, we checked  
that after an initial transient time, for \ROO=7 and 17, $\I(\tau)$  
grows exponentially but with a reduced exponent $r$ well approximated by:
\begin{equation}
\label{eqnrate}
r = \frac{1}{4} \left( 1 + p + \sqrt{1+10 p -7 p^2} \right) \ROOenfla -1
\end{equation}

Assuming for the moment that $\I(\tau)$ presents this behavior 
 also for low \ROO\ values, 
the condition for exponential growth $r>0$ leads to:

\begin{equation}
\ROOenfla > \frac{4}{1 + p + \sqrt{1+10 p -7 p^2}}  \label{eqncond}
\end{equation}

Since the right side of (\ref {eqncond})  is a strictly decreasing 
function of $p$ that takes all the values of the interval [1,2],
the condition \ROO$>1$ does not guarantee that inequality  
(\ref{eqncond}) holds.

In Fig. 1 we plot 
the curves $r=0$ and $\ROd=1$ 
in the $(p,\ROOenfla)$ plane. 
While the curve for $r=0$ is  explicitly obtained from (\ref{eqnrate}),
the curve for $\ROd=1$ is  computed numerically from (\ref{eqnr0det}) 
by sweeping the $(p,\ROOenfla)$ plane.
These curves divide the plane into three regions. In principle, one would expect exponential
epidemic growth only in region III. We denote $p_e$ the value of $p$
that satisfies $r=0$ for a given \ROO. From (\ref{eqnrate}), $p_e$
is well approximated by the value:
\begin{equation}
p_e = \frac{\ROOenfla +1 - \sqrt{(\ROOenfla+7)(\ROOenfla-1)}} {2 \ROOenfla} \label{eqnpe}
\end{equation}

\subsubsection{Numerical results}

We first show that after an initial transient time,
$\I(\tau)$ behaves as exp($-r \tau$) 
with $r$ given by expression  (\ref {eqnrate}) 
also for low \ROO\ values, 
and that therefore Fig.1 contains useful 
information. For  this purpose we first keep \ROO=1.2 fixed and 
compute $\NI(\tau)$
for different values of $p$ crossing the different regions 
of the phase diagram of Fig.1.

The results are shown in Fig. 2a and 2b. For $p$ lower than $p_e \simeq 0.383$, the 
$\NI(\tau)$-curves fall down
after having reached a maximum with a value barely greater than 
1 individual (Fig.2a). For $p$=0.383, $\NI(\tau)$ remains almost constant at the scale 
of Fig.2a, for $\tau > 5$. For all $p$-values the curves show an exponential
behavior after an initial transitory regime. It is remarkable that 
there is none qualitative change in the $\NI(\tau)$ behavior for
$p$=0.176 ($\ROd \simeq 1$). The subsequent dynamics of $\NI(\tau)$  
is shown in Fig.2b. For $p>p_e$ ($r>0$) the classical epidemic behavior is observed.
It is remarkable that for $r=0$, $\NI(\tau)$ decreases slightly from 1
individual after 800 infectious periods. It means that
the DA predicts that even when an epidemic does not develop,
the disease may persist for a very long time.
The total number of individuals that get infected in the population
 is the asymptotic number of recovered individuals, $\NR(\infty)$.
In Fig.2c it is plotted 
as a function of $p$ for \ROO=1.2, 1.4 and 1.6. The curves show a sharp
increase for $p=p_e$, giving support to the idea that  in the
phase diagram of Fig.1 region III corresponds to a system behavior qualitatively different
than in regions I and II, not only for the dynamic evolution of $\NI(\tau)$, 
but also for the 
total number of infections produced in the population.
    
\subsection{Stochastic model predictions}

\subsubsection{Averaged magnitudes. Comparison between deterministic 
and stochastic approaches}

It is not obvious that the DA equations are useful to approach the 
dynamics of the stochastic model. For high values of \ROO, in our previous work 
\cite{maltz} we checked 
that the number of infected individuals averaged over several samples was 
well approximated by the DA during the epidemic spread. Moreover, 
the fraction of infected individuals
 for each single epidemic sample  was almost identical to each other
except for a shift in time that was caused by the different instants 
at which the epidemic was triggered (see Fig.1 in \cite{maltz}). 
Of course, 
there was always a probability that extinction would occur before 
the infected individual can infect anyone, and a fraction of the samples 
did not lead to an epidemic.

For low values of \ROO, the situation for individual samples is quite different.
Almost half of the samples (48\%) become extinct within the infectious period ($\tau=1$),
but the samples that spread for longer times present a very different dynamical
behavior.
Then, in this case,  with the DA we could {\it at best} obtain a good approximation 
for the average behavior of the SM. 

We now present some results showing that the DA captures the essential behavior 
of the averaged magnitudes obtained with the SM.
In Fig.3a and 3b the dynamical behavior of 
the averaged number of infected individuals for \ROO=1.2 and different values
of $p$ is shown for the SM.
As can be observed in the figure, 
when $p$ is changed, the time evolution of  $\langle N_I \rangle$ in the SM (Fig.3a and 3b)  
presents a similar behavior to that obtained with the DA for $\NI$ (Fig.2a and 2b). 
In particular, 
it can be seen that there should be a value, $p_e '$, for $p$ 
(which corresponds to the $p_e$ of the DA)
such that $\langle N_I \rangle$  grows  exponentially with time for $p>p_e '$. 
To determine $p_e '$ we analyze  $\langle N_I \rangle$ for $\tau \in (5, 30)$ 
and check whether it is approximately constant in this interval. 
For \ROO=1.2 a value $p_e ' \simeq 0.401 \gtrsim p_e \simeq 0.383$ 
is  obtained (see Fig.3a). 
The magnitude of $\langle N_I\rangle $  is, 
however, strongly overestimated by $\NI(\tau)$ computed with the DA because of extinctions. 

The values of $p_e '$ for other \ROO\ values (determined as described above) 
are compared with the deterministic values $p_e$ in Table 1.  
It can be seen that $p_e ' \gtrsim p_e$ where
the difference $p_e ' - p_e$ increases with \ROO.
The value $R_0(p_e ')$ is also listed in the table 
 to visualize how much $R_0$ deviates from  the
 value of 1 when exponential growth has already been prevented.
$R_0$ is very well approximated by the DA in the whole range of $p$ values 
for \ROO=1.2.
For larger \ROO\ the agreement worsens but even for \ROO=1.6
the relative agreement between SM and DA computations of $R_0$ is within 3\%.

\begin{table}
\caption{
Threshold value for the global contact parameter, $p$, 
above which exponential epidemic spread is expected
in the DA ($p_e$) and the SM ($p_e '$) for different \ROO\ values. 
The value of $p_e$ is given by (\ref {eqnpe}), while
the value of $p_e '$ is determined numerically (see text).
$R_0(p_e ')$ is the basic reproductive ratio for the SM
at  $p=p_e '$. 
}

\begin{tabular}{llll}
 \\
                        \hline
                        \hline
 \ROO    &  $p_e$ & $p_e '$ & $R_0(p_e ')$ \\
                        \hline
 1.2     & 0.383  & 0.401  & 1.08   \\
 1.4     & 0.202  & 0.237  & 1.17   \\
 1.6     & 0.103  & 0.151  & 1.25   \\
 2.0     & 0.000  & 0.0672  & 1.39   \\
                        \hline
                        \hline
\end{tabular}
\end{table}

Fig.3c shows the total number of infected cases until  extinction averaged
over several samples in the SM for different values of $p$ and \ROO.
As can be observed in the figure, a sudden increase of this magnitude 
occurs      for $p \sim p_e '$,  
as was previously observed in Fig.2c for the DA results. In 
 the SM, however,  the transition  is not as
sharp as the one observed for the DA.
$\NR(\infty)$ also strongly overestimates $\langle N_R(\tau_{ext})\rangle$.
For \ROO=1.2 and $p=1$, for example, $\NR(\infty) \simeq 200,770$ 
while $\langle N_R(\tau_{ext})\rangle \simeq  33,326$.
We may conclude that the DA
predicts the correct trend for $\langle N_R(\tau_{ext})\rangle$
 and its qualitative change of behavior 
when the proportion of global contacts, $p$, is increased.
The origin of the quantitative discrepancy between $\NR(\infty)$
and $\langle N_R(\tau_{ext})\rangle$ will be studied in the following 
section considering the contribution
of each single sample to the last average.

\subsubsection{Final size of epidemics.}

The stochastic model has an inherent probabilistic nature and, 
therefore, so do its predictions.
When the infected individual enters the fully susceptible population, 
an  epidemic (of a given size) may be unleashed, or there may be no epidemic.
Average magnitudes are useful analysis tools
              but they do not have an epidemiological correlate.
In order to study the behavior of individual samples, and how it
changes  as a function of the global contact parameter, $p$, 
in Fig.4 we plot  the value of the final size of each epidemic 
as a function of the time it lasts for a set of 100,000 samples.

The figure shows that for $p \lesssim p_e '$ there is a strong correlation
between the time elapsed until extinction and the number of infections
produced during the epidemic spread (Fig. 4a-4c).
The monotonous increase of $\langle N_R(\tau_{ext})\rangle$ with $p$ 
observed in this $p$-range (Fig.3c)
could be associated with the increasing number of samples
that last longer.
For values of $p$ distinctively larger than $p_e '$ (Fig.4e and 4f)
a change of behavior is observed: 
  the points of the figures are grouped into two clearly 
  differentiated sets. The set that groups the higher values of $N_R$
forms a sort of cloud that is clearly separated from the rest of the points
that are distributed similarly to   the cases corresponding to  $p < p_e '$. 
The points in the cloud have to be identified with the samples that experienced
exponential epidemic spread, while the other set includes the ones
that became extinct without developing a major outbreak. 
For the case $p=1$, for example, if  $\langle N_R(\tau_{ext})\rangle$
is computed only considering the points in the cloud, a value of
around 200,799 individuals is obtained, which is very close to that 
predicted by the DA (200,770). This confirms that the main failure of the DA 
is not accounting for extinctions.
Finally, for $p \gtrsim p_e '$ (Fig.4c and 4d) there is a transition region 
where an intermediate behavior is observed between those described for low
and high values of $p$. In particular, for $p=0.45$ (Fig.4d),
the grouping of the points for high $N_R$ begins to be noticeable, 
but the gap in $N_R$-values observed in Fig.4e and 4f is absent yet.
 In Fig.5a we present the histograms
with the probability $G_n$ of obtaining an outbreak of
size $N_R(\tau_{ext})=n$ for the cases considered in Fig.4.
The curves quantify the changes in the distribution of outbreak sizes around $p_e '$,
observed qualitatively in Fig.4. In particular, for  $p=p_e '$, a power law behavior 
for $G_n \propto n^{-3/2}$ is observed for $10 \lesssim n \lesssim 2000$. 
The same behavior has been reported for the $G_n$ corresponding 
to the classical stochastic SIR 
model when \ROO=1 \cite{bennaim2004,bennaim2012}.

\subsubsection{System behavior at and near the threshold. Comparison with the SIR model.}
\label{sectionthreshold}
The stochastic SIR model presents a threshold for epidemic spreading at
 $\ROOenfla \equiv \beta/\gamma$=1 that has been studied by other authors
\cite{martinLof1998,bennaim2004,bennaim2012,kessler2007}.
For the SM we defined the threshold from the equation
\begin{equation}
r(p, \ROOenfla) =0 \label{eqnrsm}
\end{equation}
where $r$ is the exponent obtained by fitting  $\langle N_I \rangle$ to $k$.exp($r \tau$) for $\tau \in (5, 30)$. 
Thus, for each value of \ROO\ there is a value of $p=p_e '$, that defines the threshold: $(p_e', \alpha)$.
We computed $G_n$ for the SM, for different ($p_e '$, \ROO), and obtained
the same behavior in all cases (Fig.5b). 
 The asympthotic behavior of $G_n$ for an infinite system (solid line)
 is also indicated in the figure. By comparison,
it can be inferred that for the finite systems 
 the deviation of  $G_n$ from the asymptotic behavior 
is due to finite size effects. These effects are expected because the power law
behavior of $G_n$ implies that the outbreaks at the threshold have a probability
of reaching any size, and this is limited by the size of the system.
 
In ref. \cite{bennaim2004,kessler2007,bennaim2012} the authors studied the size effects for 
the SIR model at the threshold and argue that the maximal size of an outbreak should scale as $N^{2/3}$. 
To check their hypothesis they computed the probability $U_n(N)$ of having an outbreak
of size greater than $n$ in a system of size $N$, and found that the curves
$U_n(N)/U_n(\infty)$ vs $n/N^{2/3}$ collapse for different values of $N$. 
To see whether the SM verifies the same scaling law we  performed simulations
for the cases \ROO=1.4,  $p=0.237$ and   \ROO=1,  $p=1$ (SIR-model) considering three system sizes:
$N=N_0=640,000$, $N=2N_0$, and $N=4N_0$. 
The collapse of the curves, shown in Fig.6, indicate that the SM and the SIR model
obey the same scaling law. However, the collapse curve for \ROO=1.4 is different
from that for \ROO=1.

In the case of the SIR model, the asymptotic behavior of $G_n$ as  $n^{-3/2}$ implies that,
for an infinite system, the average size of an outbreak, $\langle n \rangle= \sum n G_n$ diverges.
However, below the threshold, it is well known 
\cite{bennaim2004,bennaim2012,kessler2007} 
that 
\begin{equation}  
\langle n \rangle = \frac{1}{1-\ROOenfla} \label{eqnnmed}
\end{equation}  
and a finite number of average infected individuals 
is expected for \ROO\ $< 1$. 
For a finite system of size $N$, if $1/(1-\ROOenfla) \ll N$, one would expect
eqn.(\ref{eqnnmed}) to hold even for a finite system \cite{kessler2007}.

In the case of the SM studied in the present work we do not have an expression
equivalent to eqn.(\ref{eqnnmed}). In order to explore the behavior of 
$\langle N_R(\tau_{ext})\rangle$ with the "distance" to the threshold, 
in Fig.7a we replot the information of Fig.3c for the cases where $p<p_e'$
taking $-1/r$ as the independent variable. The figure shows that
$\langle N_R(\tau_{ext})\rangle$ follows a linear behavior with $-1/r$,
as in the case of the SIR model, but with a slightly increasing slope 
for increasing \ROO. In Fig.7b we compare the behavior 
of $\langle N_R(\tau_{ext})\rangle$ when approaching the threshold:
\ROO=1.4, $p_e'=0.237$, by  decreasing $p_e'$ (keeping \ROO=1.4) or
decreasing \ROO\ (keeping $p_e'=0.237$). In both cases a linear behavior with
similar values for the slope of the straight lines  was obtained.
Of course, if the threshold is approached such that $-1/r$ increases beyond the scale of the figure,
 the points depart from the linear behavior due to finite size effects.

Finally, concerning the goodness of the DA to approach the SM behavior, 
it is worth mentioning that the case $p=0$ (where only local contacts 
are present) was considered by Souza  {\it et al.} \cite{souza} both 
for the SM and the DA. The authors found that the SM presents a 
percolation-like transition for \ROO\ $\simeq$ 4.6657, 
while their pair-wise approximation (our DA) predicts a transition 
at  \ROO\ =2. It is worth noting that the consideration of global 
contacts drastically improves the performance of the DA as 
can be observed in Table 1.

\section{Conclusions}

In this work we studied a stochastic epidemiological SIR model
with local and global contacts where the weight of global contacts
is given by a parameter $p$. 
Taking appropriate units of time, 
there is only another free parameter in this model, 
the quotient between transmission and recovery rates: \ROO.

    By using a deterministic approximation of the model, 
we were able to construct a phase diagram in the 
($p$, \ROO) plane where we identified a region
in which exponential epidemic spread is prevented even 
when the basic reproductive ratio of the model,  $R_0$, 
remains above 1.
 We obtained an analytical expression $p_e$(\ROO) that approximates, 
for each \ROO, the threshold value of $p$
for the exponential growth of the number of infected individuals.

We found that these predictions of the DA are closely linked
to the behavior of the average number of infected individuals
 in the SM, $\langle N_I \rangle$. In this case, we 
could define a threshold
value $p_e '$ (which is slightly larger than $p_e$)
such that  $\langle N_I \rangle$
does not grow exponentially with time for $p<p_e '$.
The absolute values of the average number of infected individuals
and the average epidemic size do not match  the
corresponding magnitudes obtained by the DA because of extinctions.
For $p>>p_e '$ the agreement could be recovered if the average
is performed using the samples that lead to epidemics exceeding
a threshold that, for high $p$ values, is well defined.

We found that the SM behavior around the threshold 
is closely related to the behavior of the classic stochastic SIR model 
around its threshold.
We summarize below the similarities and differences between both behaviors
that emerge from the comparison of our results for the SM
and what is known from the literature for the standard SIR model 
\cite{bennaim2004,bennaim2012,martinLof1998,kessler2007}.

\begin{itemize}
\item[-] 
In the SIR model, the threshold is given by $\beta / \gamma \equiv \ROOenfla$=1. 
As $R_0$=\ROO\ and
the exponent for epidemic spreading is $r=\ROOenfla-1$, the threshold may be expressed
unambiguously as $R_0=1$ or $r=0$.

In the SM both $R_0$ and $r$ are functions of \ROO\ and $p$. As is well known
(and was verified in the present work)
if the population is not well mixed $R_0$ is not a useful concept \cite{anderson,failurer0,riley}.
The threshold is obtained from the condition $r=0$, which in this case does not give a 
parameter value, but a relation between them:
$r(p,\ROOenfla) = 0$ that leads to $p=p_e'(\ROOenfla)$. 

\item[-] 
Below the threshold, the average size of an  outbreak in the SIR model verifies:
\begin{equation}
\langle N_R(\tau_{ext})\rangle=\frac{1}{1-\ROOenfla} = - \frac{1}{r}
\end{equation}

For the SM we empirically obtained that $\langle N_R(\tau_{ext})\rangle$ grows approximately linearly
with $-1/r$. However, if $p$ or \ROO\ is kept fixed, the behavior is strictly linear:
\begin{equation}
\langle N_R(\tau_{ext})\rangle=  \frac{a(\ROOenfla)}{r} + b(\ROOenfla) = \frac{c(p)}{r} + d(p)
\end{equation}
with $a$, $b$, $c$ and $d$ smooth functions.
In both cases the above relationships are no longer valid in the 
vicinity of the threshold due to finite size effects.

\item[-] 
Strictly at the threshold, the probability of having an outbreak 
of size $n$ in the SIR model is $G_n =k n^{-3/2}$ for $n \in (n_1,n_2)$, where
$n_2$ increases with the size of the system, $N$.

In the SM the same behavior is observed for all the $(p_e',\ROOenfla)$
cases considered.

\item[-] 
The probability $U_n(N)$ that the outbreak size is at least $n$ 
in a population of size $N$ shows the following scaling law
for the SIR model \cite{bennaim2004,bennaim2012}.
\begin{equation*}
\frac{U_n(N)}{U_n(\infty)}=f\left( n/N^{2/3}\right)
\end{equation*}

The same scaling law was observed in the present work
for the SM, where the scaling function
$f$ depends on the threshold  point 
$\left( p_e'(\ROOenfla), \ROOenfla \right)$, $f$ being
an increasing function of \ROO.
\end{itemize}

It would be interesting to know if those properties at the threshold, which the SM shares with the stochastic SIR model, 
hold for other networks with global and local contacts

Our study highlights the importance of keeping the global contacts
as low as possible as a key measure to prevent large epidemics
and points out that a substantial improvement of the epidemiological 
status (where exponential epidemic spread is prevented)
could be accompanied by an insignificant reduction of $R_0$,
which remains with values well above one.
Even though the SM model is very simple, from the epidemiological
perspective and in the treatment of the spatial structure,
we believe that our conclusions could be taken as a basis for
 exploration by more complex models in specific contexts.

\section{Acknowledgments}
This work was supported by Agencia Nacional de Promoci\'on
     Cient\'{\i}fica y Tecnol\'ogica-ANCPyT grant PICT2010-0707
and Universidad Nacional de La Plata grant X805(2018-2019). G.F. is
member of the Scientific Career of Consejo Nacional de Investigaciones 
Cient\'{\i}ficas y Tecnol\'ogicas-CONICET (Argentina).

\section{Figure captions}

\vskip 1cm
Figure 1

Phase diagram for  epidemic growth in the deterministic 
approximation. The regions above and below the dashed line correspond
 to the points ($p$, \ROO) with $\ROd$ greater and lower than 1 respectively.
 The regions above and below the solid line correspond
 to the rate of epidemic growth, $r$, greater and lower than 0 respectively.

\vskip 1cm
Figure 2

DA results.
(a) Number of infected individuals ($\NI$) as a function of time ($\tau$) for
\ROO=1.2 and $p$=0.10, 0.176, 0.30, 0.35, 0.383, 0.40 and 0.45. The values 
$p$=0.176 and 0.383 correspond to $\ROd \simeq 1$ and $r\simeq 0$ respectively. 
(b) $\NI$ for \ROO=1.2 on an expanded time scale and 
$p$=0.10, 0.30, 0.35, 0.383, 0.40, 0.45, 0.50, 0.60 and $1.0$.
The solid line corresponds to $p=0.383\simeq p_e$.
(c) Asymptotic number of infected individuals $\NR(\infty)$ as a function of $p$ for three \ROO\
values. Vertical lines indicate the $p_e$ values: 
0.383, 0.202, 0.103 corresponding to
\ROO=1.2, 1.4, and 1.6 respectively.

\vskip 1cm
Figure 3

Average magnitudes in the SM.
(a) Averaged number of infected individuals, $\langle N_I \rangle$,
for \ROO=1.2 and different values of $p$. The thin horizontal line
has been drawn to show that $\langle N_I (\tau) \rangle$ for $p=0.401$
remains almost constant for $\tau \in (5,30)$.
(b) $\langle N_I \rangle$ for \ROO=1.2 on an expanded time scale and 
for $p=0.10, 0.30, 0.35, 0.401, 0.45, 0.50, 0.60$ and $1.0$. 
The solid line corresponds to $p=0.401\simeq p_e '$.
Each curve is the average of $m$ independent stochastic simulations,
where $m=100,000$ for $p\geq 0.50$ and $m=1,000,000$ for $p < 0.50$.
(c) Averaged number of individuals that have experienced the infection 
until  extinction for different values
of \ROO\ and $p$. Each point corresponds to an average over 
100,000 samples when $\langle N_R(\tau_{ext})\rangle > 5000$, 
and to an average over 
1,000,000 samples when $\langle N_R(\tau_{ext})\rangle < 5000$.
The crosses correspond to $p=0.401\simeq p_e '(1.2)$, 
$p=0.237\simeq p_e '(1.4)$, and $p=0.151\simeq p_e '(1.6)$.  
The broken lines have been drawn to guide the eye.
The vertical solid      lines indicate the $p_e$ values corresponding
to the beginning of exponential spread in the DA, as in Fig.2c.

\vskip 1cm
Figure 4

Number of individuals that have experienced the infection
in a given stochastic simulation, $N_R(\tau_{ext})$, as a function
of the duration of the corresponding simulation, $\tau_{ext}$, for \ROO=1.2. Different
panels correspond to simulations performed for different $p$ values.
Each panel contains 100,000 points, each one corresponding to an independent
stochastic simulation with identical initial conditions.

\vskip 1cm
Figure 5

Probability $G_n$ of obtaining an outbreak of size $N_R(\tau_{ext})=n$. 
 (a) $G_n$ for \ROO=1.2 and different $p$ values. 
 The different curves represent the distributions of points in the $y$-axis 
 of the different pannels of Fig.4. 
 (b)  $G_n$ at the threshold for different cases.
  Black line:
  $n^{-3/2} / \sqrt{4\pi}$ 
  (asymptotic behavior 
  for an infinite system in the case $p=1$, $\alpha=1$
    \cite{bennaim2012}).

\vskip 1cm
Figure 6

Scaling behavior of the normalized cumulative distribution
$U_n(N)/U_n(\infty)$  at the threshold.
Six different systems, corresponding to two sets of parameters at the
threshold ($p(\alpha),\alpha$), and three system sizes ($N$), were considered.
When plotted as a function of $n/N^{2/3}$ the distributions
corresponding to $\alpha=1.4$  or $\alpha=1$ colapse in the upper or
lower curves, respectively. $N_0=640,000$.

\vskip 1cm
Figure 7

Averaged number of individuals that have experienced the infection 
until  extinction below the threshold.
(a) The cases \ROO=1.2, 1.4, and 1.6 for different values of $p$ are considered. 
The points correspond to the same simulations
of Fig.3c (cases with $p<p_e '$) but $\langle N_R(\tau_{ext}) \rangle$
is plotted here versus $-1/r$, where 
 $r$ is determined
fitting $\langle N_I (\tau) \rangle$ to exp($r \tau$) for $\tau \in (5,30)$
as in Fig.3a.
The broken lines are linear fits to the points
corresponding to a fixed \ROO. The fitted slopes are 
 1.154, 1.357, and 1.669 for \ROO=1.2, 1.4 and 1.6 respectively.
The unit slope line (solid) is the expected behavior for the SIR model
(\ROO=1, $p=1$).
(b) The case \ROO=1.4 for different values of $p$, and
    the case $p=0.237$ for different values of \ROO\ are considered.
   Broken lines are linear fits corresponding to each case with slopes
    1.357 ($\alpha=1.4$) and  1.382 ($p=0.237$).
Note that the two cases correspond to two perpendicular directions
on approaching the threshold ($p=0.237$, \ROO=1.4) in the ($p,\alpha$)-plane.

\newpage

{\large Figure 1}
\vskip 0.5cm

\includegraphics[width=10cm,height=7cm]{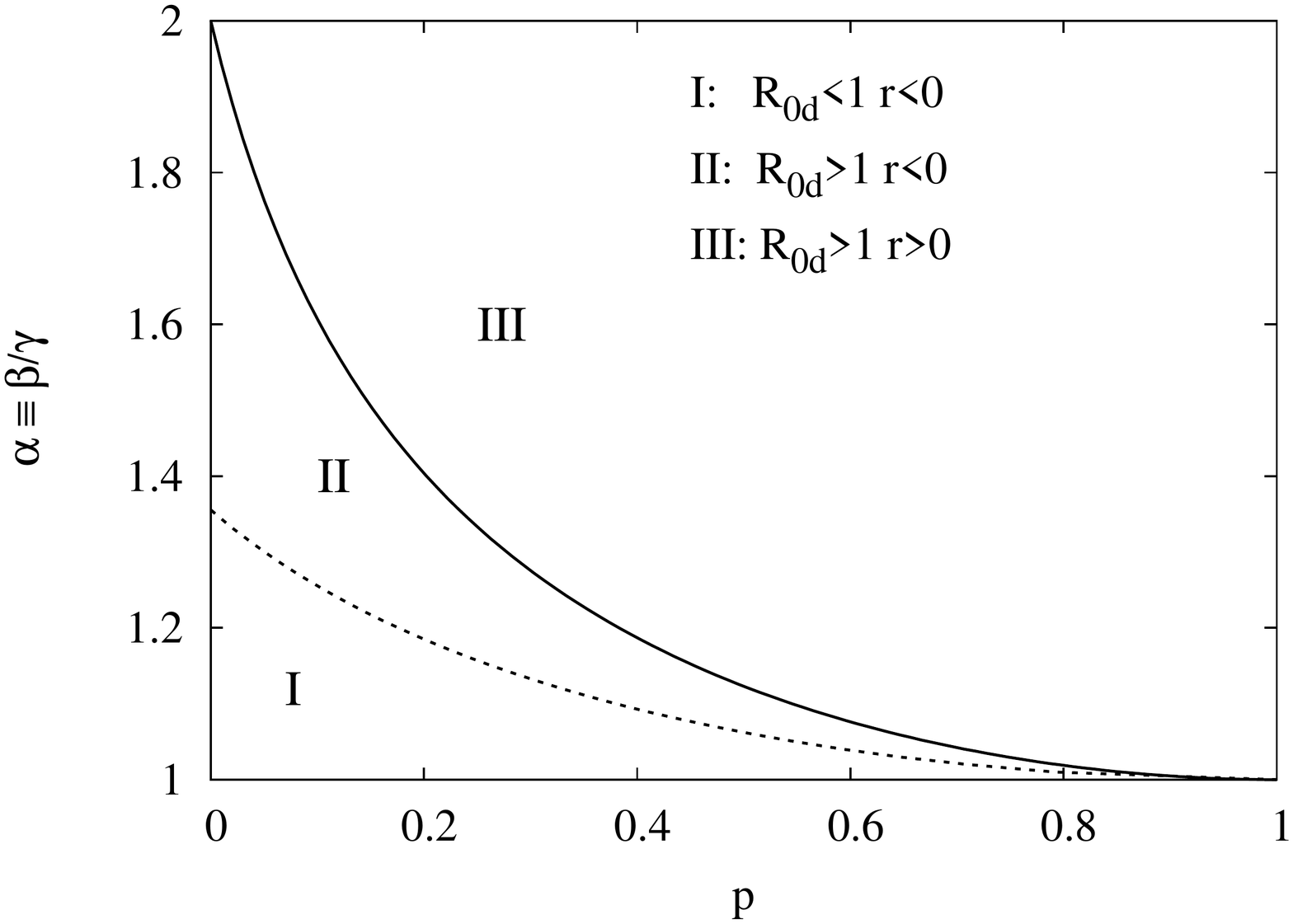}
\vskip 1cm

\newpage
\newgeometry{left=3cm,top=1cm} 
{\large Figure 2} 

\vskip 2cm
{\large (a)} 
\vskip -2cm
\hskip  3cm
\includegraphics[width=10cm,height=7cm]{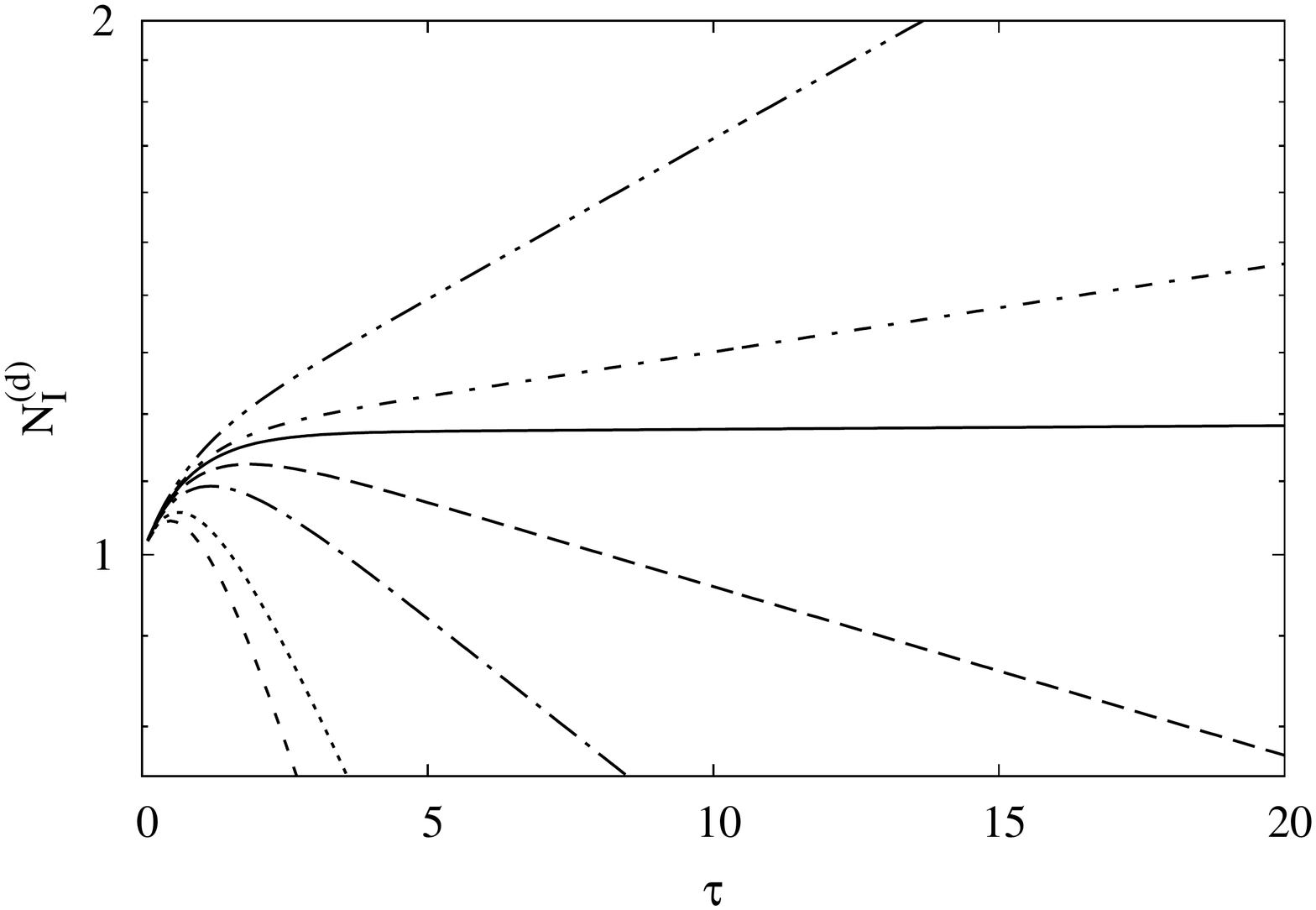}

\vskip 2cm
{\large (b)} 
\vskip -2cm
\hskip  3cm
\includegraphics[width=10cm,height=7cm]{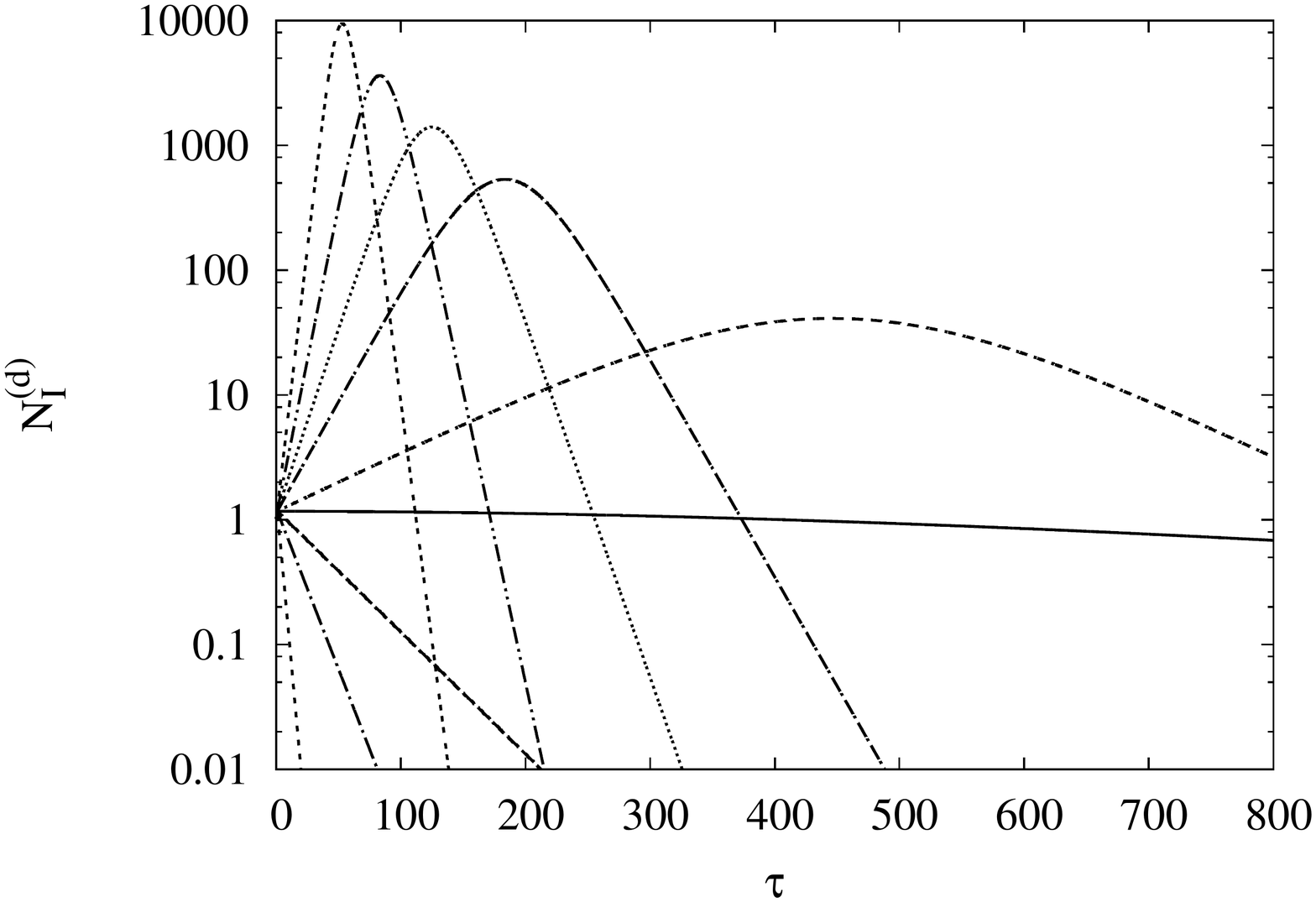}

\vskip 2cm
{\large (c)} 
\vskip -2cm
\hskip  3cm
\includegraphics[width=10cm,height=7cm]{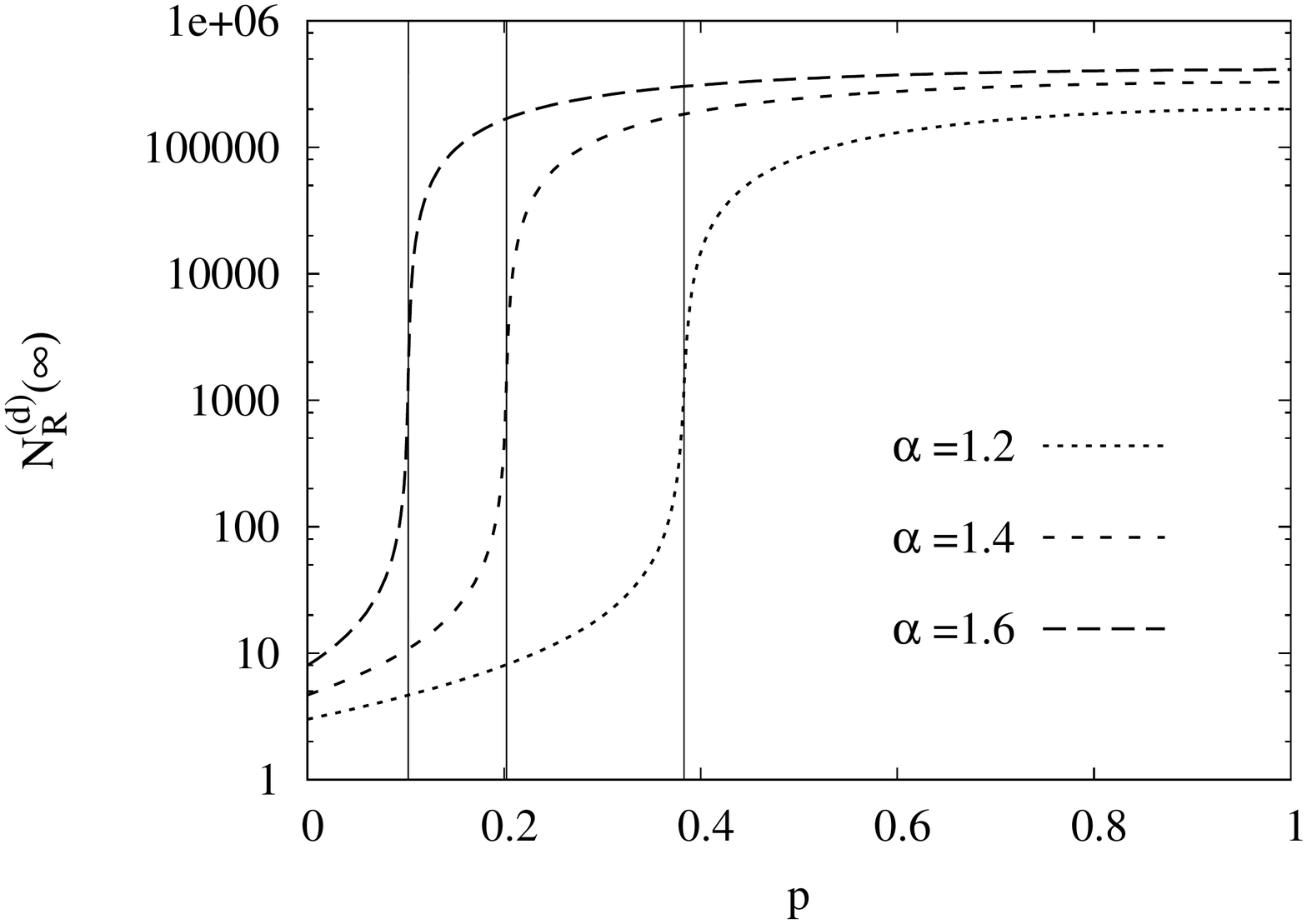}
\restoregeometry

\newpage
\newgeometry{left=3cm,top=1cm} 
{\large Figure 3} 

\vskip 2cm
{\large (a)} 
\vskip -2cm
\hskip  3cm
\includegraphics[width=10cm,height=7cm]{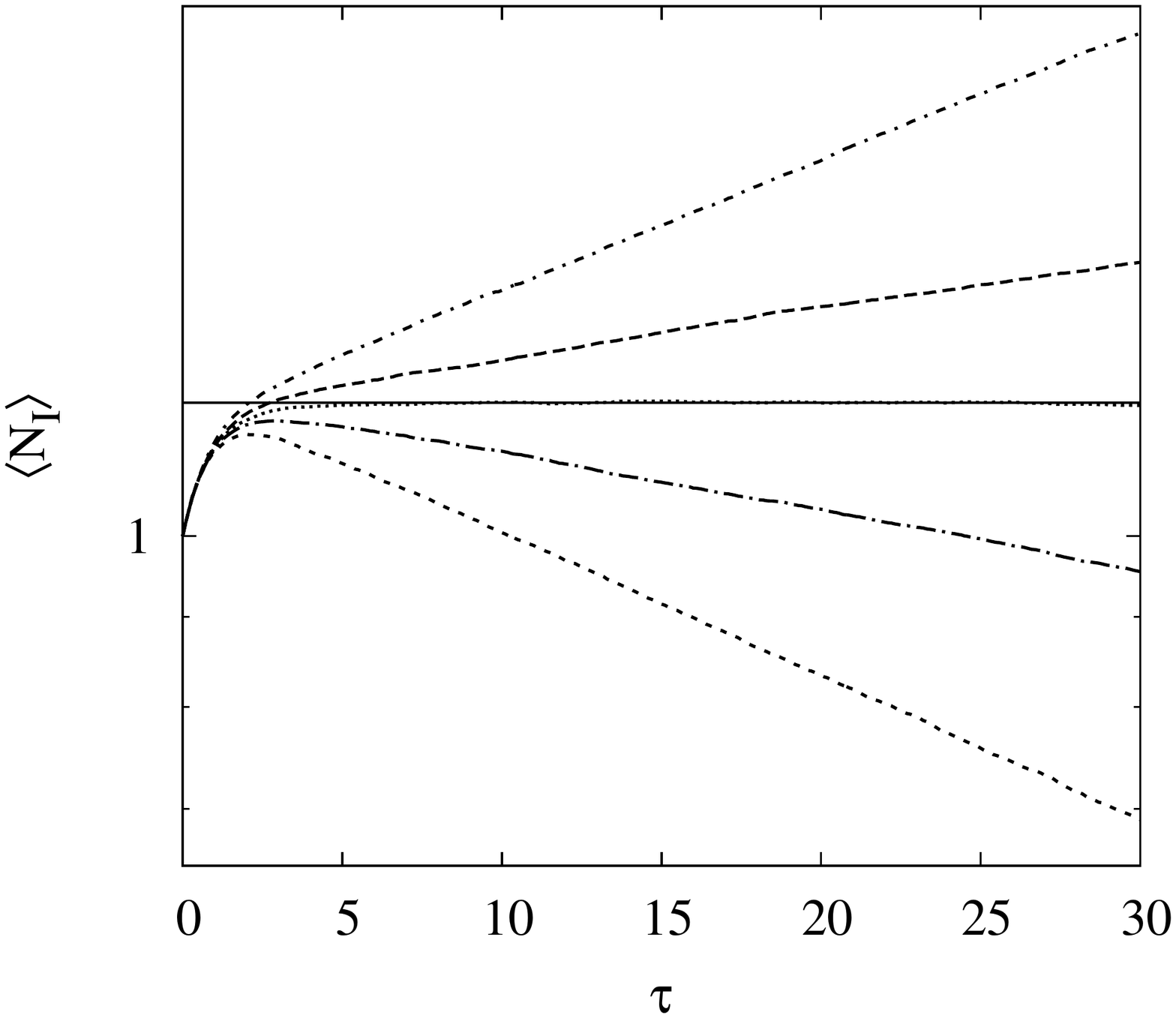}

\vskip 2cm
{\large (b)} 
\vskip -2cm
\hskip  3cm
\includegraphics[width=10cm,height=7cm]{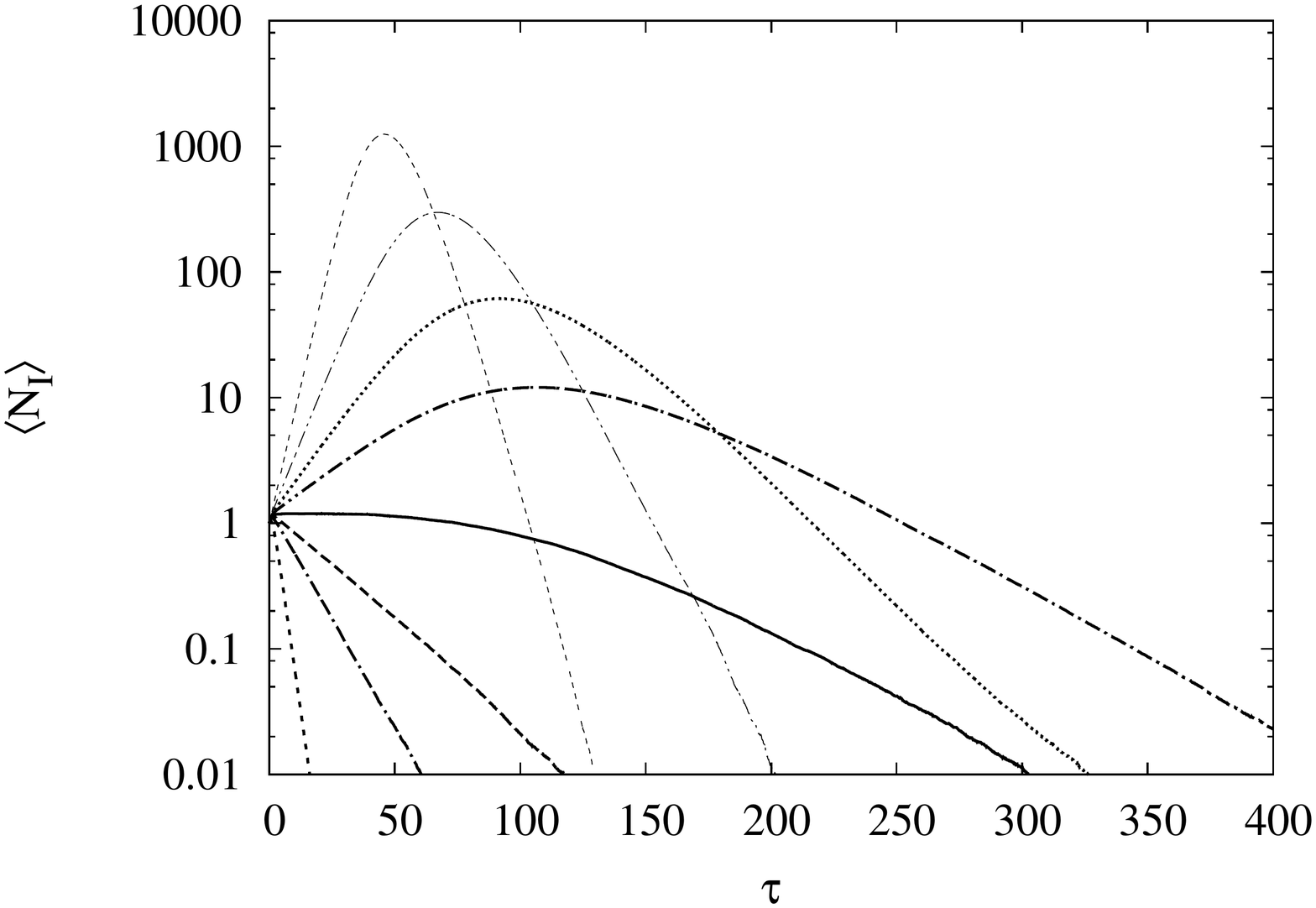}

\vskip 2cm
{\large (c)} 
\vskip -2cm
\hskip  3cm
\includegraphics[width=10cm,height=7cm]{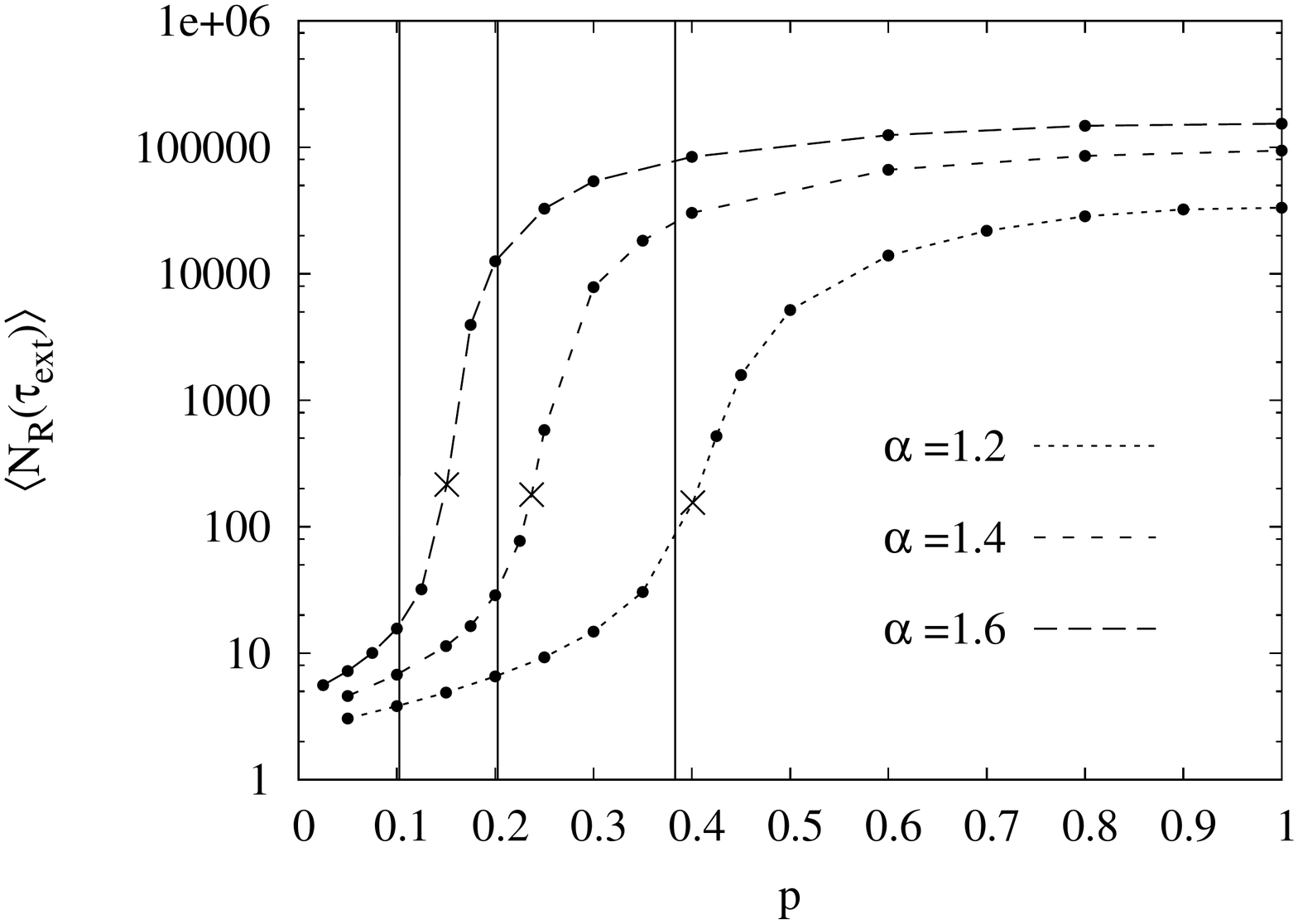}
\restoregeometry

\newpage
{\large Figure 4a} \hskip 8cm {\large Figure 4b}
\vskip 0.5cm

\hskip -2cm
\includegraphics[width=10cm,height=7cm]{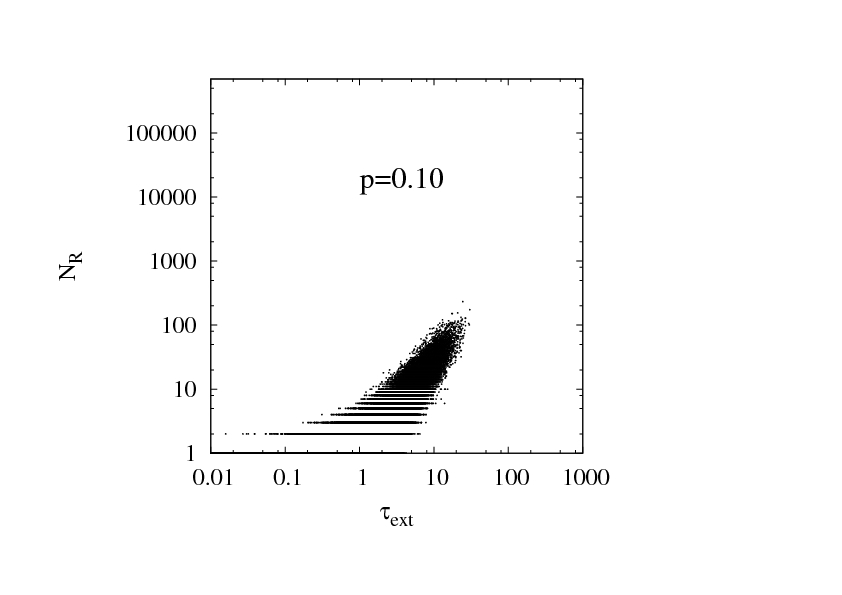}
\includegraphics[width=10cm,height=7cm]{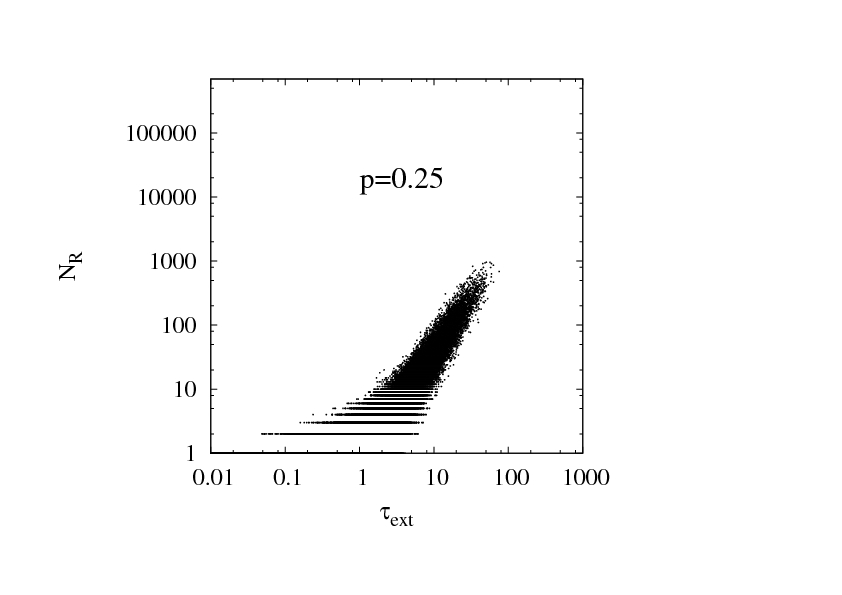}
\vskip 1cm

{\large Figure 4c} \hskip 8cm {\large Figure 4d}
\vskip 0.5cm

\hskip -2cm
\includegraphics[width=10cm,height=7cm]{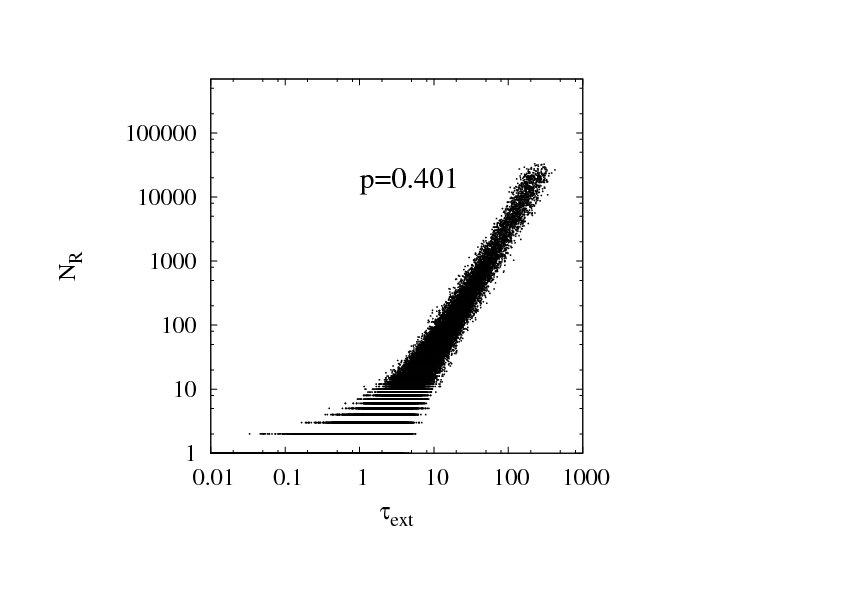}
\includegraphics[width=10cm,height=7cm]{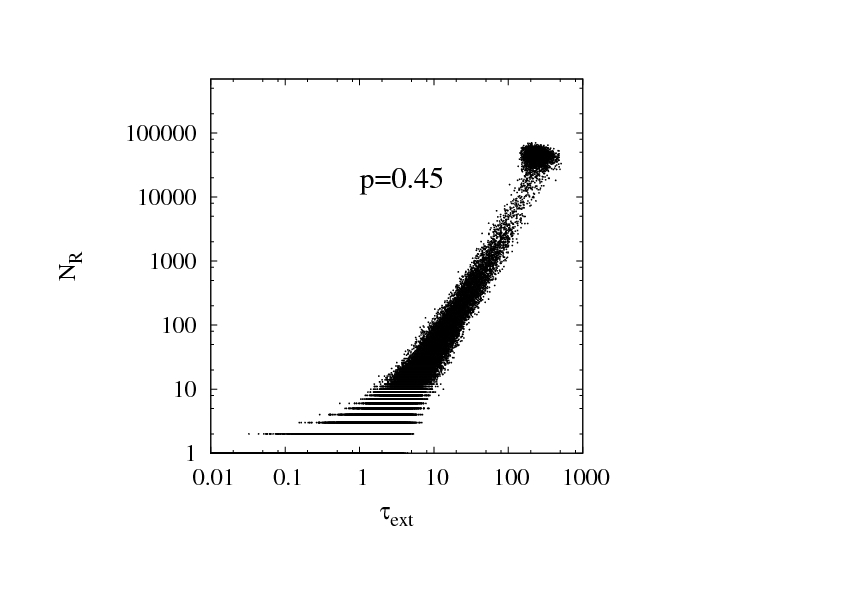}
\vskip 1cm

\newpage
{\large Figure 4e} \hskip 8cm {\large Figure 4f}
\vskip 0.5cm

\hskip -2cm
\includegraphics[width=10cm,height=7cm]{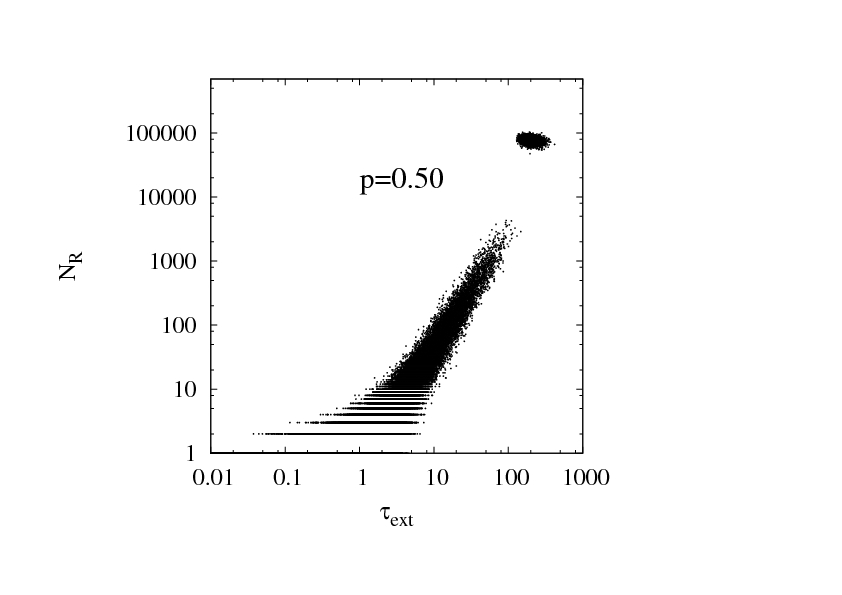}
\includegraphics[width=10cm,height=7cm]{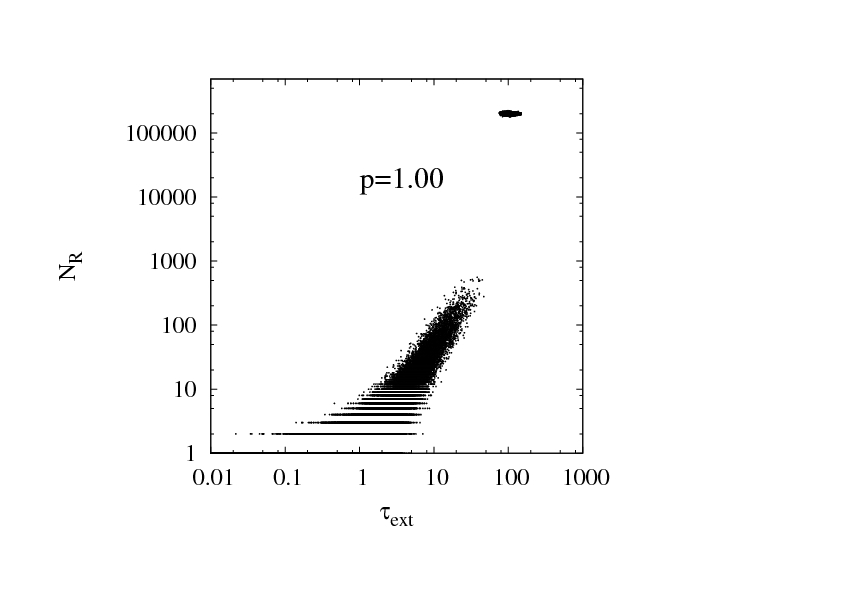}
\vskip 1cm

\newpage
{\large Figure 5a} \hskip 8cm {\large Figure 5b}
\vskip 0.5cm

\hskip -2cm
\includegraphics[width=9cm,height=7cm]{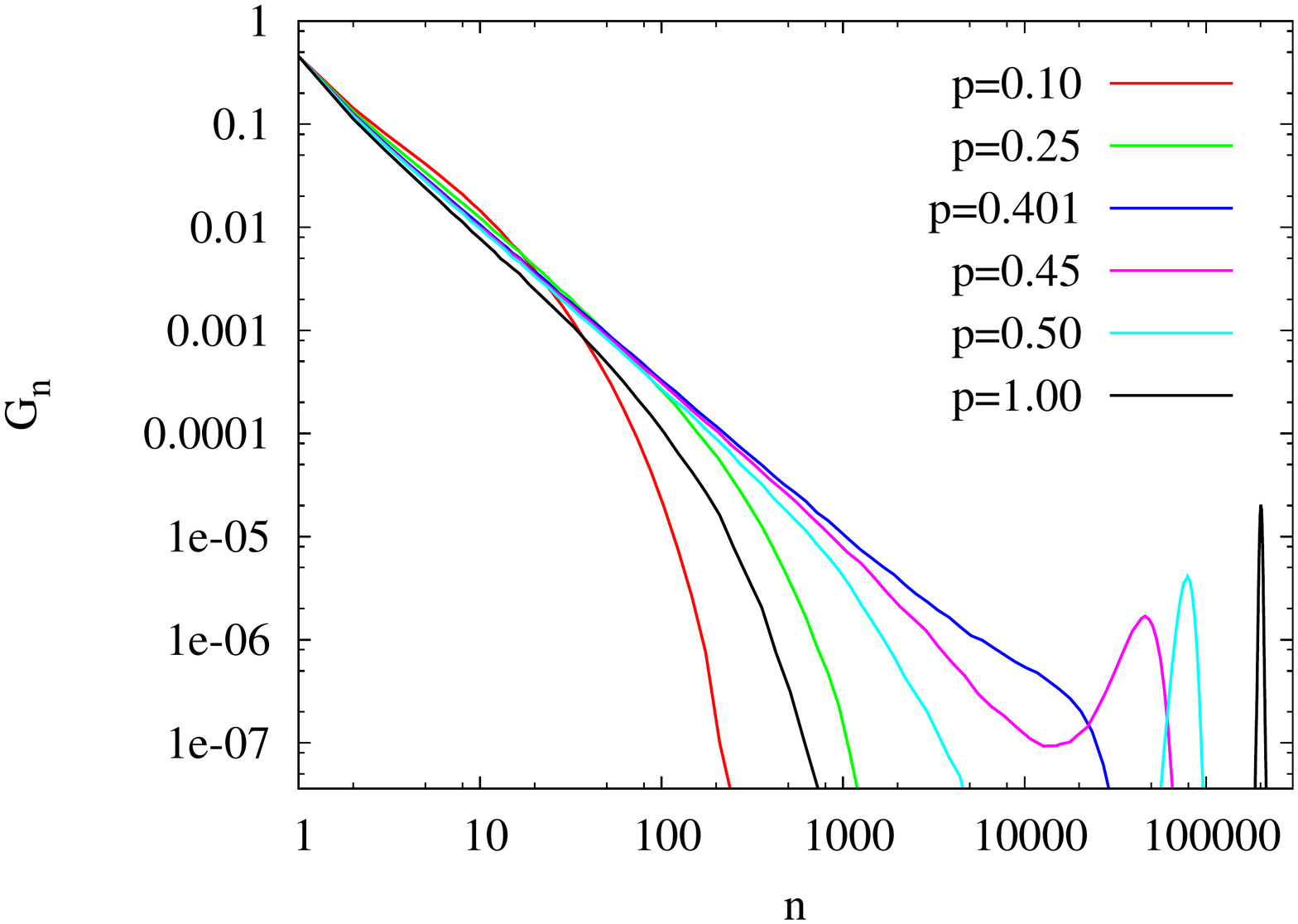}
\includegraphics[width=9cm,height=7cm]{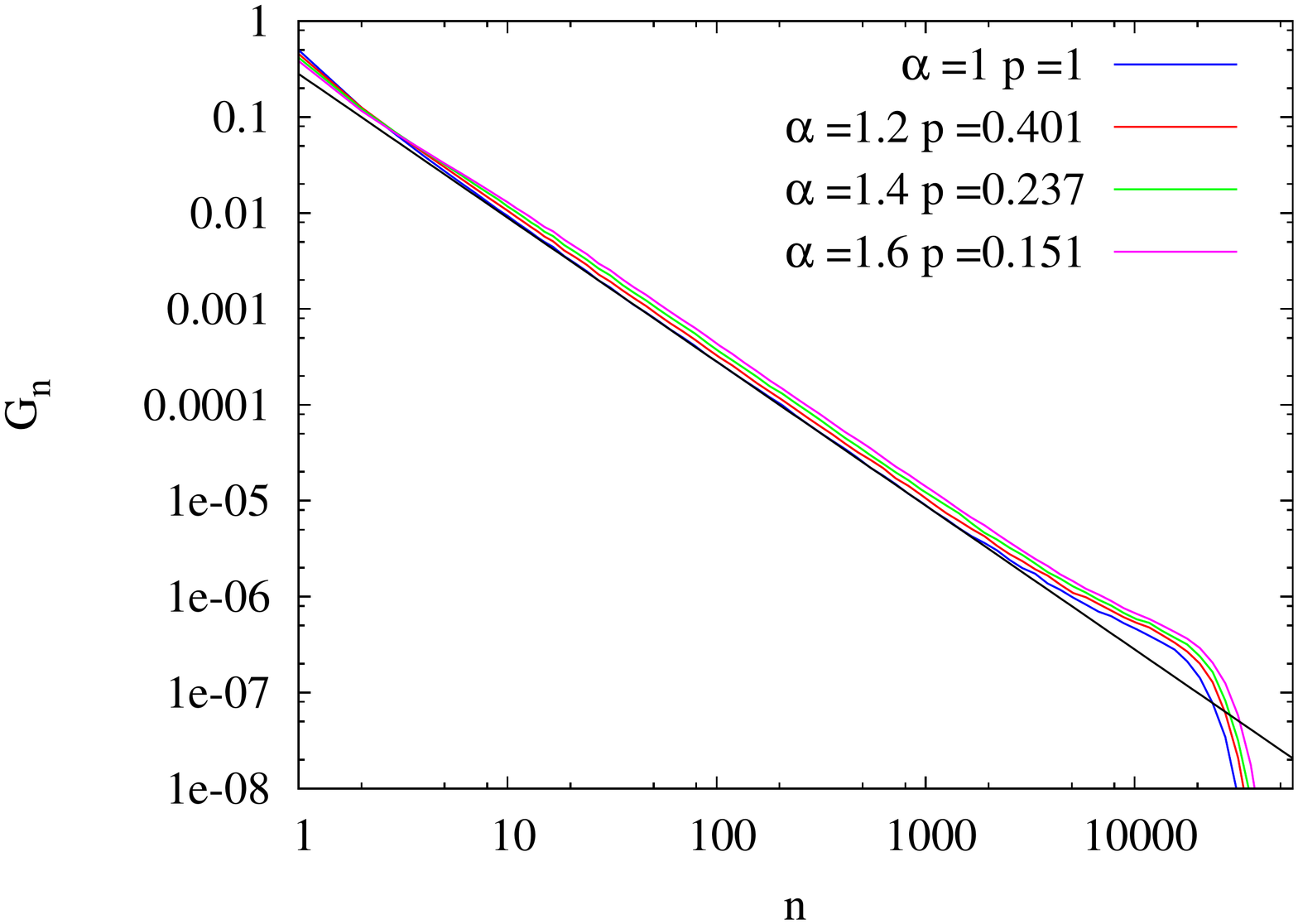}
\vskip 1cm


{\large Figure 6}
\vskip 0.5cm

\includegraphics[width=10cm,height=7cm]{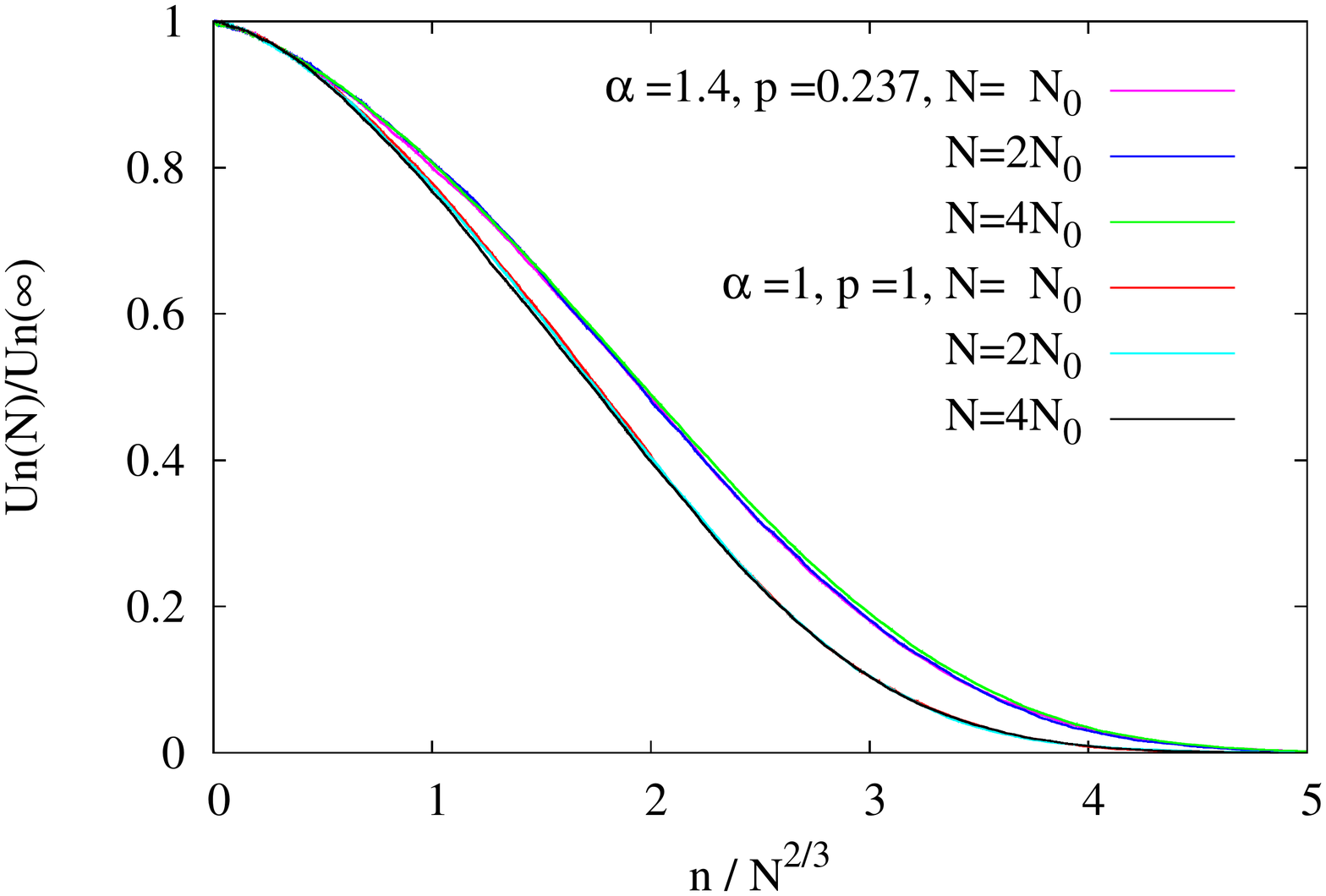}
\vskip 1cm

\newpage
{\large Figure 7a} \hskip 8cm {\large Figure 7b}
\vskip 0.5cm

\hskip -2cm
\includegraphics[width=9cm,height=7cm]{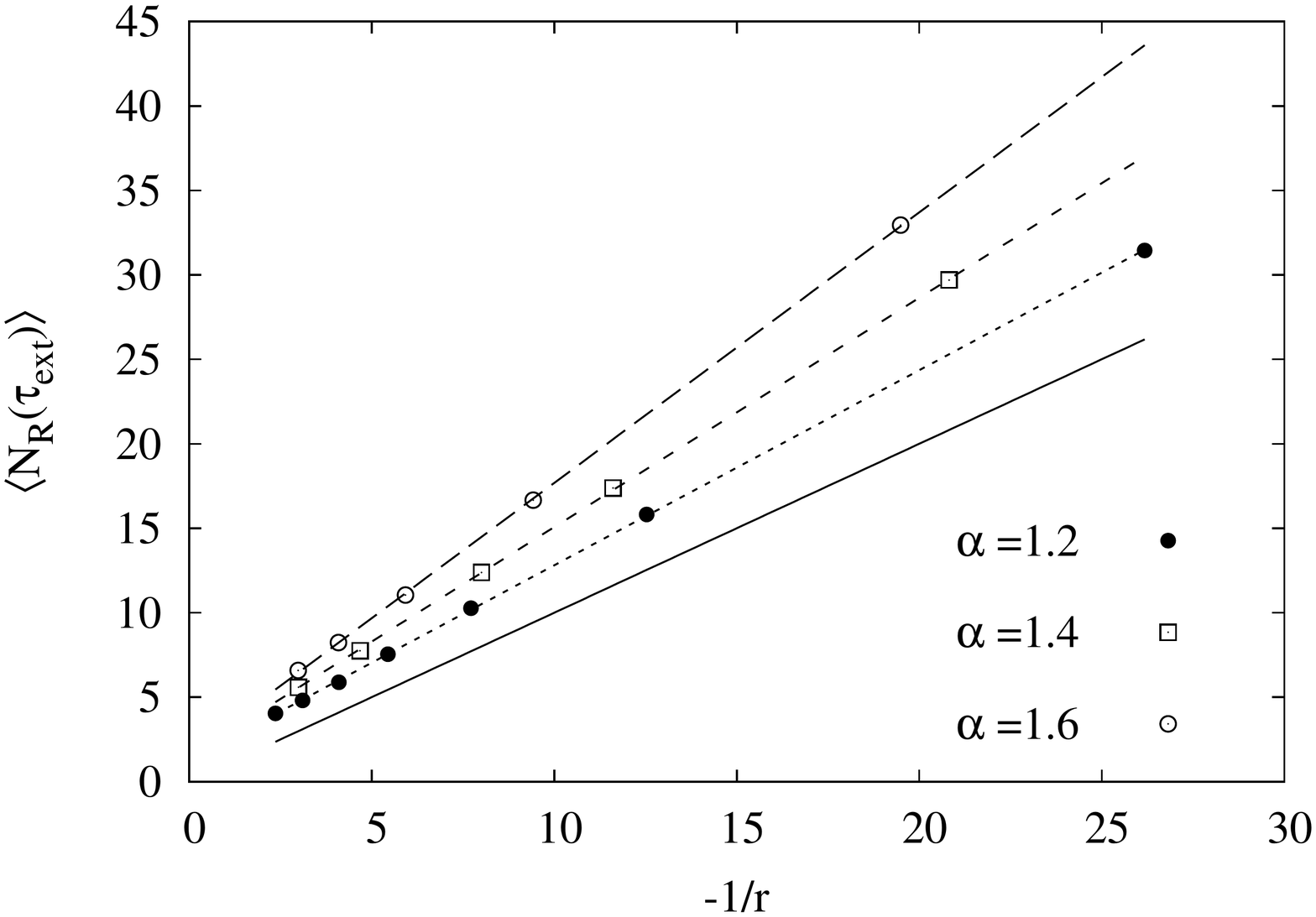}
\includegraphics[width=9cm,height=7cm]{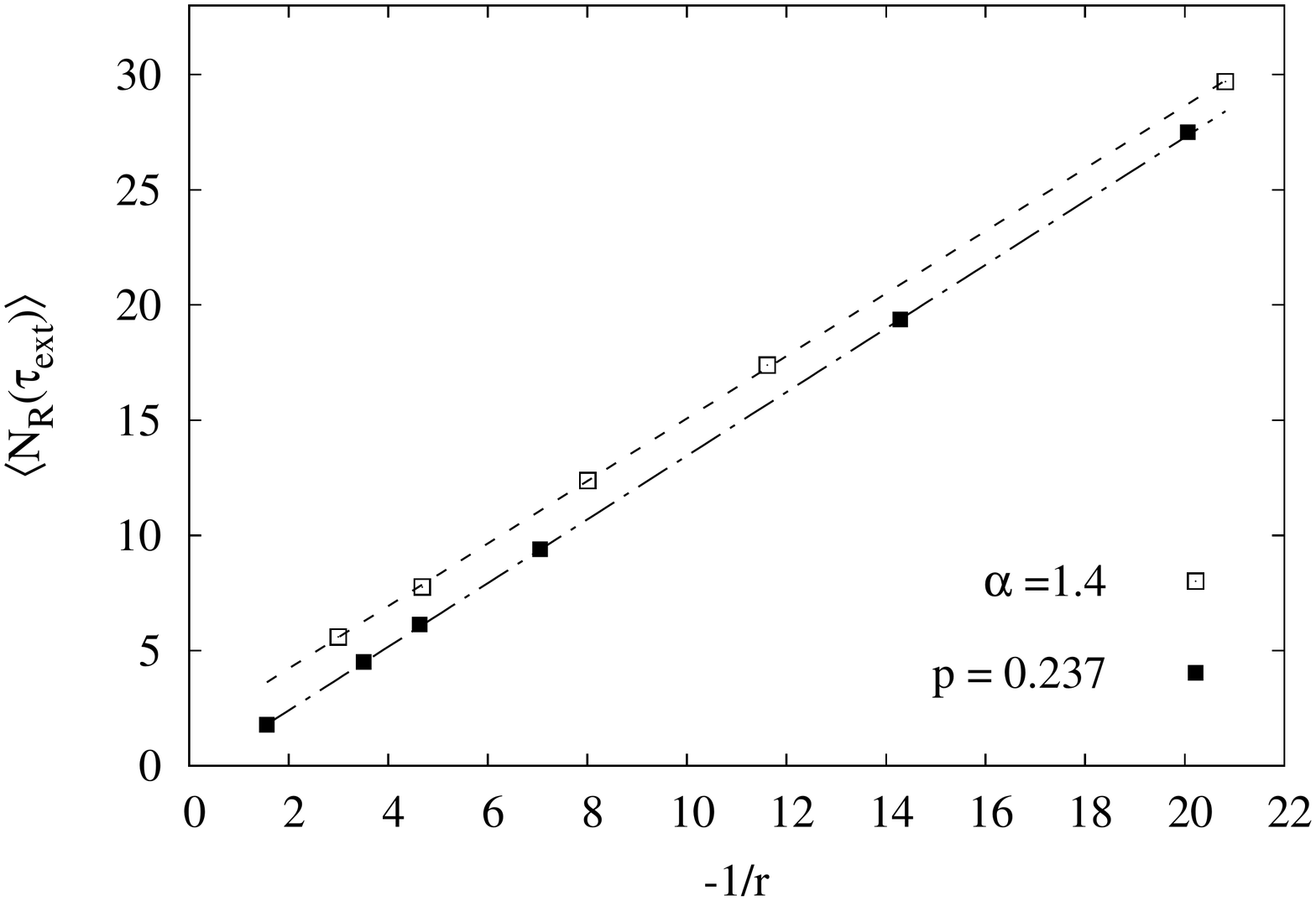}


\newpage


\section*{References}

\begin{thebibliography} {00}

\bibitem{bailey} N.Bailey

  {\bf The Mathematical Theory of Infectious Diseases and its Applications}
   
Charles Griffin and Company, London, 1975

 \bibitem{anderson} R. Anderson and R. May

   {\bf Infectious Diseases of Humans: Dynamics and Control }

   Oxford University Press, Oxford, 1991

\bibitem{libroKR} M.Keeling and P.Rohani

{\bf Modeling infectious diseases in humans and animals}

 Princeton University Press, 2008

 \bibitem{kermack} W. Kermack and A. McKendrick

   {\bf A contribution to the mathematical theory of epidemics}

    Proc.Roy.Soc. Lon. A  {\bf 115} (1927) pp. 700-721

\bibitem{heester2}  
    J. Heesterbeek, R. Anderson, V. Andreasen, S. Bansal, D. DeAngelis, C. Dye,
  K. Eames, W. Edmunds. D. Frost, S. Funk, T. Hollingworth, T. House, V. Isham,
P. Klepac, J. Lessler, J.Lloyd-Smith, C. Metcalf, D. Mollison, L. Pellis,  J. Pulliam,
 M. Roberts, and C. Viboud                  
 
 {\bf  Modelling infectious disease dynamics in the complex landscape of global
 health}
 
 Science    {\bf 347}  (2015)
 
 DOI: 10.1126/science.aaa4339

\bibitem{daley}  D.Daley and D.Kendall

{\bf  Epidemics and rumours}

 Nature {\bf 204} (1964)  pp. 1118-1118

\bibitem{mishra} B.Mishra and D.Saini

{\bf Mathematical models on computer viruses}

 Applied Mathematics and Computation  {\bf 187}  (2017)  pp. 929-936
 
\bibitem{woo}J. Woo and H.Chen

{\bf Epidemic model for information diffusion in web forums: experiments in marketing
exchange and political dialog}
 
SpringerPlus {\bf 5} (2016)

DOI : 10.1186/s40064-016-1675-x  

\bibitem{shive} S.Shive

{\bf An epidemic model of investor behavior}

 Journal of financial and quantitative analysis   {\bf 45} (2010)  pp. 169-198


   \bibitem{bartlett56} M. S. Bartlett

   {\bf Deterministic and stochastic models for recurrent epidemics}


    Proc. Third Berkeley Symp. on Math. Statist. and Prob.

     
    Univ. of Calif, Press  {\bf 4}  (1956) pp. 81-109   

 \bibitem{bartlett57} M.S. Bartlett

   {\bf  Measles Periodicity and Community Size}

Journal of the Royal Statistical Society. Series A (General)   {\bf 120}  (1957)  pp. 48-70

 \bibitem{martinLof1998}  A. Martin L\"of
  
      {\bf The final size of a nearly critical epidemic and first passage 
     time of a Wiener process to a parabolic barrier}

       J.App.Prob. {\bf 35}  (1998)  pp. 671-682
      
    \bibitem{bennaim2004}   E. Ben Naim and P. Krapivsky
   
    {\bf Size of outbreaks near the epidemic threshold}

    Phys.Rev E  {\bf 69} (2004)

    DOI: 1103/PhysRevE.69.050901
     
   \bibitem{kessler2007} D.Kessler and N.Shnerb

      {\bf Solution of an infection model near threshold}

      Phys.Rev E {\bf 76} (2007)  010901(R)


    \bibitem{bennaim2012} E. Ben Naim and P. Krapivsky

     {\bf Scaling behavior of threshold epidemics}

     The Eur.Phys.Journ. B   {\bf 85} (2012)

     DOI: 0.1140/epjb/e2012-30117-0

    
\bibitem{watts-strogatz} D.Watts and S.Strogatz 

{\bf Collective dynamics of `small world' networks}

Nature  {\bf 393} (1998)  pp. 440-442

\bibitem{verdasca} J. Verdasca, M. Telo da Gama, A. Nunes,
N. Bernardino,  J. Pacheco and M. Gomes

 {\bf Recurrent epidemics in small world networks }

  J. Theoret. Biol.  {\bf 233}  (2005)  pp. 553-561   
\bibitem{simoes}   M. Sim\"oes, M. Telo da Gama and A. Nunes

     {\bf Stochastic fluctuations in epidemics on networks}
     
     J.R. Soc. Interface  {\bf 5}  (2008) pp. 555-566
 \bibitem{dottori} M. Dottori and G. Fabricius

  {\bf SIR model on a dynamical network and the endemic state of
  an infection disease}
  
   Physica A  {\bf 434}  (2015) pp. 25-35.  
  \bibitem{maltz}  A. Maltz and G. Fabricius

{\bf  SIR model with local and global infective contacts: a deterministic
  approach and applications }

    Theoret.  Popul. Biol  {\bf 112}  (2016)  pp. 70-79

  \bibitem{gilligan} C.Gilligan
  
  {\bf Sustainable agriculture and plant diseases: an epidemiological perspective}
  
  Phil.Trans R.Soc B {\bf 363} (2008)  pp.741-759


 \bibitem{liccardo} A.Fierro, A.Liccardo and F.Porcelli

 {\bf A lattice model to manage the vector an the infection of
  the Xyllela fastidiosa on olive trees}

   Nature, Scientific Reports   {\bf 9} (2019) Article number 8723
  
  \bibitem{tischendorf} L.Tischendorf and C.Staunbach
  
  {\bf Chance and risk of controlling rabies in large-scale and long-term
  	inmunized fox populations}
  
  Proc.Roy.Soc. B {\bf 265} (1998)  pp. 839-846
     
\bibitem{failurer0} J.Li, D. Blakeley and R.Smith

{\bf The failure of R0}

Comput.Math.Methods.Med.  (2011)   2011.527610

\bibitem{riley} S. Riley, K. Eames, V. Isham, D. Mollison and P. Trapman

  {\bf Five challenges for spatial epidemic models }

  Epidemics {\bf 10}  (2015)  pp. 68-71

\bibitem{gillespie} D. Gillespie

   {\bf A general method for numerically simulating the stochastic
 time evolution
 of coupled chemical reactions}
 
   J. of Comput.  Phys. {\bf 22}  (1976)  pp. 403-434   
\bibitem{heester1} J. Heesterbeek and K. Dietz

  {\bf The concept of  $R_0$ in epidemic theory }

    Statistica Neerlandica   {\bf 50}  (1996)  pp.89-110

\bibitem{souza} D. R. Souza and T. Tom\'e

{\bf Stochastic lattice gas model describing the dynamics of the SIRS
epidemic process}

Physica A  {\bf 389}  (2010) 1142-1150
                   
\end{thebibliography}
\end{document}